\documentclass[12pt]{article}
\newcommand{\be}{\begin{equation}}
\newcommand{\ee}{\end{equation}}

\begin{document}

\begin{center}
\vspace*{1.5in}
{\Large \bf Nonleading charmed quark mass corrections} \\
\vspace{0.2in}
{\Large \bf to $K^0-\bar{K^0}$ mixing in the standard model
\footnote{Electronic version of the published paper, 
the journal reference: Phys.~Lett.~B263~(1991)~282}}\\
\vspace{0.6in} {\large \bf A.A.~Pivovarov} \\
\vspace{0.1in} {Institute
for Nuclear Research of the Russian Academy of Sciences\\ Moscow
117312} \\ 
\vspace{0.5in}
{\bf Abstract} 
\end{center}
\thispagestyle{empty}

The nonleading corrections in the inverse mass of the heavy charmed quark to
the effective local $\Delta S=2$ Lagrangian of $K^0-\bar{K}^0$ mixing are
calculated within the standard model.

\newpage
1. Nowadays the  standard model of strong and electroweak interactions [1-3]
is the subject of intensive and thorough experimental study \cite{4}. So far the
model goes successfully through the numerous precious tests even though the
accuracy of the experimental data is such that the radiative corrections must
be properly taken into account  \cite{5}. One of the most interesting effects having
got the explanation within the standard model is the phenomenon of
CP-invariance breaking observed firstly in the system of the neutral kaon
but later found for the $B^0-\bar B^0$ as well. The problem of CP-invariance
breaking is still very intriguing and both experimental and theoretical
investigations of the $K^0-\bar{K}^0$ system draw much attention. In this
particular case, however, the direct comparison of the theoretical predictions
with experimental data is made somewhat difficult due to the necessity of
determination of the hadronic matrix elements of the effective electroweak
Lagrangian. It is this problem which causes the main part of uncertainties of
the theoretical calculations. The effective electroweak local $\Delta S=2$
Lagrangian for the $K^0-\bar{K}^0$ mixing in the leading approximation in the
heavy charmed quark mass $m_c$ has been obtained in the famous paper by
Gaillard and Lee  \cite{6}. Then the leading perturbation theory 
corrections due to strong
interactions were calculated in  \cite{7}. The corresponding hadronic matrix element
of the effective local $L_{\Delta S=2}$
Lagrangian between $K^0$ and $\bar{K}^0$ states
\[
\langle \bar K^0(k')|L_{\Delta S=2}|K^0(k)\rangle
\]
has been intensively studied during several last years with different
techniques [8-17] and, probably, the final result is consistent with the
vacuum saturation hypothesis  \cite{6} within $20\%$ accuracy.

However, Wolfenstein  \cite{18} has pointed out that the local
effective Hamiltonian
does not exhaust the whole physics of $\Delta S=2$ transitions. It cannot
account for the long distance contribution which is present in the initial
Green's function for the matrix element of the $K^0-\bar{K}^0$ mixing and is
connected with the propagation of the light $u-$quark round the loop of the
box diagram. This contribution is purely non-perturbative and ultimately
depends on the infrared properties of QCD. There are some model dependent
estimates of this part of the entire matrix element in the literature [19-23]
though the accuracy is still far from being satisfactory.

Nevertheless, there is one point which can be essentially improved
just within perturbation theory for the standard model. It consists in the
calculation of the corrections in the inverse mass of the charm quark to the
local part of the effective Lagrangian. These correction can be important
because the charmed quark is not heavy enough in comparison with the
characteristic scale in the sector of light $(u,d)$ quarks, for example, with
the $\rho-$meson mass, $m_\rho$. The analogous corrections to the decays of
charmed baryons and mesons have been recently considered in refs. [24-27].

In the present paper the non-leading corrections in $m_c^{-1}$ to the local part
of the effective $\Delta S=2$ Lagrangian are calculated. They are represented by
the local operators with dimension eight in mass units.

2. The local effective  $\Delta S=1$  
Lagrangian after decoupling of the W-boson has the form
\be
\label{one}
L_{\Delta S=1}={G_F\over \sqrt2}J_\mu J_\mu^{+}
\ee
where $G_F$ is the Fermi constant, $J_\mu=\bar{Q}_L\gamma_{\mu}Vq_L$ is
the weak charged hadronic current, $Q=(u,c,t)^T$, $q=(d,s,b)^T$, $q_L$
denotes the left handed Dirac spinor, and $V$ is the quark mixing matrix.
Then the matrix element $M$ of the $K^0-\bar{K}^0$ transition is
represented by the expression
\[
_{out}\langle \bar K^0(k')|K^0(k)\rangle_{in}=i(2\pi)^4\delta (k-k')M,
\]
\be
\label{two}
M={i\over2}\int \langle \bar K^0(k')|TL_{\Delta S=1}(x) 
L_{\Delta S=1}(0)|K^0(k)\rangle dx. 
\ee

Strictly speaking, formulae (\ref{one}) and (\ref{two}) 
are literally valid for the
situation where the $t-$quark is much lighter than the W-boson. As is well
known now this is not the case anymore. For our purpose, however, it is not
essential and in the following we neglect the $t-$quark mixing at all
and restrict ourselves to the simplified model with two generations only.

The effective  $\Delta S=2$ Lagrangian can be written in the form
\be
\label{three}
L_{\Delta S=2}=\left(\frac{4G_F \sin\theta_c \cos\theta_c}{\sqrt2}\right)^2(L_H+L_L)
\ee
where $\theta_c$ is the Cabibbo angle and we have used the obvious
normalization. Here
\be
\label{threea}
L_H=i\int T_H(x)dx,~~T_H=T_{cc}-T_{cu}-T_{uc} 
\ee
is the heavy part of the whole effective  $\Delta S=2$ Lagrangian containing
the loops with a virtual heavy $c-$quark in the intermediate state, while
\be
\label{threeb}
L_L=i\int T_L(x)dx,~~T_L=T_{uu} 
\ee
describes the light part of the transition.
We also introduce the useful notations
\[
T_{uu}(x)=T\bar{s}_L \gamma_{\alpha}u_L \bar{u}_L\gamma_{\alpha}d_L(x)
\bar{s}_L \gamma_{\beta}u_L \bar{u}_L\gamma_{\beta}d_L(0),
\]
\[
T_{cu}(x)=T\bar{s}_L \gamma_{\alpha}u_L \bar{c}_L\gamma_{\alpha}d_L(x)
\bar{s}_L \gamma_{\beta}c_L \bar{u}_L\gamma_{\beta}d_L(0)
\]
and so on.
Expression (3) is finite due to GIM mechanism but $L_H$ and $L_L$
separately require a regularization because they are ultraviolet divergent.
The dimensional regularization is not convenient in this case due to 
the presence
of the $\gamma_5$-matrix. The Pauli-Willars regularization introduces a
regulator mass that makes it difficult to perform an explicit
calculation from the technical point of view. We will use the
regularization which is free of the shortcomings of both of them in our
particular case. Namely, let us define the regularized quantities $L^R_{H,L}$
by the equation
\be
\label{four}
L^R_{H,L}=i\int T_{H,L}(x){(-\mu^2x^2)}^{\epsilon}dx 
\ee
where $\epsilon$ is the regularization parameter and $\mu$ represents the mass
scale analogous to one of the dimensional regularization.

Now for the heavy part of the effective $\Delta S=2$ Lagrangian $L_H$ we have a
regular expansion in the inverse charmed quark mass in the following form (from
now on we omit the index ``R'')
\[
16\pi^2L_H=C_0O_0+\sum_{j}C_jO_j
\]
where $C_j$ are coefficient functions depending  on the heavy quark mass
$m_c$ and $O_j$ are  the local operators  built from the  light $(u,d,s)$ quark
field only. If we split the whole Lagrangian
into the sum
\be
\label{five}
L_H=L_H^{(0)}+L_H^{(1)} 
\ee
then
\be
\label{six}
16\pi^2L_H^{(0)}=-m_c^2(\bar{s}_L \gamma_{\alpha}d_L)^2 
\ee
is the well known result of Gaillard and Lee \cite{6}. 
The rest part of eq.~(\ref{five})
\be
\label{seven}
16\pi^2L_H^{(1)}=\sum_{j}C_jO_j 
\ee
consists of local operators which have dimension eight in mass units.
It is convenient to define the operator basis in the form
\[
O_{\tilde F} = \bar{s}_L \gamma_{\alpha}d_L \bar{s}_L
\gamma_{\mu}{\tilde F}_{\mu\alpha}d_L,
\]
\[
O_A=\bar{s}_L \gamma_{(\mu}D_{\nu)}d_L \bar{s}_L
\gamma_{(\mu}D_{\nu)}d_L,
\]
\[
O_B=\bar{s}_L
\gamma_{\mu}D_{\mu}d_L \bar{s}_L \gamma_{\mu}D_{\mu}d_L,
\]
\[
O_C= \bar{s}_L \gamma_{\alpha}d_L
\bar{s}_L(\gamma_{\mu}D_{\mu}D_{\alpha}+D_{\alpha}\gamma_{\mu}D_{\mu})d_L -
{{(m_s^2+m_d^2)}\over2} (\bar{s}_L \gamma_{\alpha}d_L)^2
\]
where
\[
\gamma_{(\mu}D_{\nu)}={1\over2}(\gamma_{\mu}D_{\nu} +\gamma_{\nu}D_{\mu}).
\]
Then a direct calculation gives
\[
16\pi^2L_H^{(1)}=-{4\over3}(O_{\tilde F}+O_A)
\left({1\over\epsilon}+\ln({4{\mu}^2e^{-2C}\over {m_c^2}})+{4\over3}\right)
-{2\over3}O_A
\]
\be
\label{eight}
-{2\over3}(O_B+O_C)
\left({1\over\epsilon}+\ln{({4\mu^2e^{-2C}\over{m_c^2}})}+{11\over6}\right)
\ee
where $C=0.577\ldots$ is the Euler constant.

Eq.~(\ref{eight}) is the main result of the present paper. After performing
a renormalization procedure 
(say, a minimal subtraction of the pole term) we will have
the finite quantity and the parameter $\mu$ 
recalls the necessity to have the proper short distance contribution 
missed in eq.~(\ref{threea}) and given by eq.~(\ref{threeb}).
At this order of the expansion in $m_c^{-1}$ the dependence on this
parameter is explicit, contrary to the leading order (\ref{six})
which is finite
and does not depend on $\mu$ at all. It is obvious from the expression
(\ref{eight}) that the
heavy and light parts of the whole Lagrangian must be defined
simultaneously in
a coordinated way. The pole part of eq.~(\ref{eight}) is canceled by the
corresponding divergences of the light part due to GIM mechanism. The last
statement can be explicitly demonstrated.

Namely, let us consider the light part. The operator product expansion for the
amplitude $T_L(x)$ from eq.~(\ref{threeb}) in $x^2$ 
at $x^2\rightarrow 0$ has the form
\be
\label{nine}
T_L(x)={1\over{4\pi^4x^6}} (\bar{s}_L \gamma_{\alpha}d_L)^2 
+{1\over{16\pi^4{x^4}}} \left({4\over3}(O_{\tilde F}+
O_A)+{2\over3}(O_B+O_C)\right). 
\ee

The short distance contribution of the light part does cancel the corresponding
divergences of the heavy part. More technically we extract the short distance
contribution of the light part as follows. We split the entire light part in the
following way
\[
L_L =i\int T_L(x){(-\mu^2x^2)}^{\epsilon}dx
\]
\be
\label{ten}
=i\int T_L(x)(f(x,\bar x)+{\bar f}(x,\bar x)){(-\mu^2x^2)}^{\epsilon}dx=
L_L^{SH}+L_L^{LG} 
\ee
where
\be
\label{tena}
L_L^{SH} =i\int T_L(x)f(x,\bar x){(-\mu^2x^2)}^{\epsilon}dx,  
\ee
\be
\label{tenb}
L_L^{LG}=i\int T_L(x){\bar f}(x,\bar x){(-\mu^2x^2)}^{\epsilon}dx, 
\ee
and
\[
f(x,\bar x)+{\bar f}(x,\bar x)=1.
\]

A simple model which is the smooth generalization of the step
$\theta-$function
is chosen for the functions $f(x,\bar x)$ and
${\bar f}(x,\bar x)$
\[
f(x,\bar x)={{\bar x^{2n}}\over{\bar x^{2n}+(-x^2)^n}},
\quad
{\bar f}(x,\bar x)={{x^{2n}\over{\bar x^{2n}+(-x^2)^n}}}
\]
in such a way that the function $f(x,\bar x)$ cuts out the short
distances only (up to $\bar x$) and the function ${\bar f}(x,\bar x)$
does the long distances.
After performing the integration we get for the short distance contribution of
the light part the expression
\[ 
16\pi^2L_L^{SH} = (\bar{s}_L \gamma_{\alpha}d_L)^2
{{\pi/n}\over \sin(\pi/n)} {4\over{\bar x^2}} +\left({4\over3}(O_{\tilde F}+
O_A)+{2\over3}(O_B+O_C)\right) 
\left({1\over\epsilon}+\ln{\mu^2\bar x^2}\right).
\]
Now the whole answer is
\[
16\pi^2(L_H+L_L)= 16\pi^2L_L^{LG}+
(\bar{s}_L\gamma_{\alpha}d_L)^2\left(-m_c^2 +{{\pi/n}\over
\sin(\pi/n)}
{4\over{\bar x^2}}\right)
\]
\be
\label{eleven}
-\left({4\over3}(O_{\tilde F}+ O_A)+ {2\over3}(O_B+O_C)\right)
\left(\ln{({4e^{-2C}\over{m_c^2\bar x^2}})}+{4\over3}\right)-{2\over3}O_A.
\ee

The long distance part of the whole light Lagrangian $L_L^{LG}$ (\ref{tenb})
cannot be calculated due to strong infrared problems and
requires some low-energy model. It can be estimated from lattice calculations
or with the help of the theory of effective chiral Lagrangians. We hope to
consider this question in a separate publication.

To get the numerical estimates for the corrections obtained in the
paper 
(eq.~(\ref{eleven})) 
we use factorization for the kaon-antikaon matrix elements of the local
operators $O_{\tilde F} - O_C$. In the chiral limit we have
\[
\langle \bar K^0(k)|(\bar{s}_L \gamma_{\alpha}d_L)^2|K^0(k)\rangle^{fact}=
(1+{1\over{N_c}})({{f_K^2m_K^2}\over2}),
\]
\[
\langle \bar K^0(k)|O_{\tilde F}|K^0(k)\rangle^{fact}
=-\delta^2 ({{f_K^2m_K^2}\over2}),
\]
\[
\langle \bar K^0(k')|O_A|K^0(k)\rangle^{fact}=
-\delta^2{1\over{N_c}}({{f_K^2m_K^2}\over2})
\]
where the parameter $\delta^2$ is defined by the relation \cite{28}
\[
\langle 0|\bar{s}_L \gamma_{\mu}{\tilde F}_{\mu\alpha}d_L|K^0(k)\rangle
=-ik_\mu f_K \delta^2
\]
and all other matrix elements vanish.

From eq.~(\ref{eleven}) we finally have
\[
16\pi^2(L_H+L_L^{SH})=
\langle\bar K^0(k)|(\bar{s}_L \gamma_{\alpha}d_L)^2|K^0(k)\rangle^{fact}
\left(-m_c^2 +{{\pi/n}\over \sin(\pi/n)}{4\over{\bar x^2}}\right.
\]
\be
\label{twelve}
\left.
+{4\over3}\delta^2 (\ln{({4e^{-2C}\over{m_c^2\bar x^2}})}
+{4\over3})+{1\over6}\delta^2\right). 
\ee

Let us estimate $\bar x^2$. From eq.~(\ref{nine}) we find
\be
\label{thirteen}
T_L(x)= {1\over{4\pi^4x^6}}
\langle \bar K^0(k)|(\bar{s}_L \gamma_{\alpha}d_L)^2|K^0(k)\rangle^{fact}
(1-{{\delta^2{x^2}}\over3}+o(x^2)). 
\ee

Then at $\delta^2\bar x^2=3$ where the expansion (\ref{thirteen}) blows up
and at $n=\infty$ we obtain the following representation for
the non-leading corrections
\be
\label{fourteen}
-m_c^2<\bar K^0(k)|(\bar{s}_L \gamma_{\alpha}d_L)^2|K^0(k)>^{fact}
(1-{4\over3}{\delta^2\over{m_c^2}}
-{4\over3}{\delta^2\over{m_c^2}} (\ln{({4\delta^2\over{3
m_c^2}})}-2C+{35\over24})). 
\ee

Numerically at $m_c^2=1.6~GeV^2$, $\delta^2=0.2~GeV^2$ \cite{28,29} we get
\[
-m_c^2\langle \bar K^0(k)|(\bar{s}_L \gamma_{\alpha}d_L)^2|K^0(k)
\rangle^{fact} (1-0.2+0.3)
\]
\[
=-m_c^2\langle\bar K^0(k)|(\bar{s}_L \gamma_{\alpha}d_L)^2|K^0(k)
\rangle^{fact} (1+0.1).
\]

But if we define the boundary value of $\bar x^2$ by the relation $\delta^2\bar x^2=1$
then the corrections are much larger and have the opposite sign
\[
-m_c^2\langle \bar K^0(k)|(\bar{s}_L \gamma_{\alpha}d_L)^2|K^0(k)
\rangle^{fact}(1-0.5+0.1)
\]
\[
-m_c^2\langle \bar K^0(k)|(\bar{s}_L \gamma_{\alpha}d_L)^2|K^0(k)
\rangle^{fact}(1-0.4).
\]

3. To conclude, we have calculated the corrections in the inverse mass of the
charmed quark, considered as heavy enough, to the effective local $\Delta S=2$
Lagrangian. In this order there already exists the explicit dependence on the
boarder between short and long distances even in the heavy part of the whole
Lagrangian. A splitting of light part into the short and long distance
contributions is proposed which respects the cancelation of purely ultraviolet
divergences, {\em i.e.} the GIM mechanism.  The OPE at $x^2\rightarrow0$
for the light part of Lagrangian is analyzed and the convergence scale is
determined explicitly. In the vacuum dominance approximation this scale is
large enough and is determined by the parameter $\delta^{-2}$. The corrections
can decrease the answer for the transition matrix element approximately twice
though it depends on the concrete choice of the scale $\bar x^2$ and also on
the validity of the vacuum dominance approximation for estimating the kaon
matrix elements of the operators $O_{\tilde F}-O_C$.  This means that the
purely long distance hadronic contribution could be important to stabilize the
whole answer in the region where the expansion (13) fails, in other words one
can predict that the integral $L_L^{LG}$ changes quickly when $\bar x^2$
belongs to the region $1/\delta^2<\bar x^2<3/\delta^2$ just to compensate the
corresponding dependence of the short distance part
(\ref{twelve}). Nevertheless,
in ordert to
determine the absolute magnitude of the long distance hadronic contribution one
needs a low-energy model such as, for example, lattice or chiral Lagrangians.


\begin{thebibliography}{99}

\bibitem{1}  S.L.Glashow, {Nucl.Phys. 22(1961)579.}
\bibitem{2}  S.Weinberg, {Phys.Rev.Lett. 19(1967)1264.}
\bibitem{3}  A.Salam, {in: Elementary Particle Theory, Ed.N.Svartholm
(Almqvist and Wiksel, 1968).}
\bibitem{4}  See, for example C.Jarlskog, {Preprint CERN-TH.5740/90,1990.}
\bibitem{5}  See, for example G.Altarelli, {Preprint CERN-TH.5834/90,1990.}
\bibitem{6}  M.K.Gaillard and B.W.Lee, {Phys.Rev. D10(1974)897.}
\bibitem{7}  F.J.Gilman and M.B.Wise, {Phys.Rev. D27(1983)1128.}
\bibitem{8}  R.E.Shrock and M.B.Wise, {Phys.Rev. D19(1979)2148.}
\bibitem{9}  J.F.Donoghue, E.Golowich and B.R.Holstein,{Phys.Lett. B119(1982)412.}
\bibitem{10} P.Colic, B.Guberina,  J.Trampetic  and  D.Tadic,
   {Nucl.Phys. B221(1983)141.}
\bibitem{11} B.Guberina, B.Machet and E.de Rafael,{Phys.Lett. B128(1983)269.}
\bibitem{12} A.Pich and E.de Rafael, {Phys.Lett. B158(1985)477.}
\bibitem{13} K.G.Chetyrkin et al.,  {Phys.Lett. B174(1986)104.}
\bibitem{14} R.Decker, {Nucl.Phys.B277(1986)661.}
\bibitem{15} L.J.Reinders and S.Yazaki, {Nucl.Phys. B288(1987)789.}
\bibitem{16} N.Bilic, C.A.Dominguez and B.Guberina, {Z.Phys. C39(1988)351.}
\bibitem{17} G.Martinelli, {Nucl.Phys. B(Proc.Suppl)16(1990)16.}
\bibitem{18} L.Wolfenstein, {Nucl.Phys. B160(1979)501.}
\bibitem{19} J.F.Donoghue, E.Golowich and B.R.Holstein,
    {Phys.Lett. B135(1984)481.}
\bibitem{20} I.I.Bigi and A.I.Sanda, {Phys.Lett. B148(1984)205.}
\bibitem{21} P.Cea and G.Nardulli, {Phys.Lett. B152(1985)251.}
\bibitem{22} B.Machet, N.F.Nasrallash and K.Shilcher,{Phys.Rev. D42(1990)118.}
\bibitem{23} Ho Tso-Lsin et al., {Phys.Rev. D42(1990)112.}
\bibitem{24} H.Georgi, B.Grinstein and M.B.Wise, {Phys.Lett. B252(1990)456.}
\bibitem{25} N.Isgur and M.B.Wise, {Nucl.Phys. B348(1991)276.}
\bibitem{26} H.Georgi, {Nucl.Phys. B348(1991)293.}
\bibitem{27} M.E.Luke, {Phys.Lett. B252(1990)447.}
\bibitem{28} V.A.Novikov et al., {Nucl.Phys. B237(1984)525;}
\bibitem{29} A.A.Ovchinnikov and A.A.Pivovarov, {Sov.J.Nucl.Phys. 48(1988)1135.}

\end{thebibliography}
\end{document}